\newcommand{\be}{\begin{equation}}
\newcommand{\ee}{\end{equation}}
\newcommand{\bea}{\begin{eqnarray}}
\newcommand{\eea}{\end{eqnarray}}
\newcommand{\bc}{\begin{center}}
\newcommand{\ec}{\end{center}}

\def\RN{Reis\-sner-Nord\-str\"{o}m }
\documentstyle[preprint,aps]{revtex}
\draft
\input{epsf}
\begin{document}
\title{The Cauchy horizon in
Higher-derivative Gravity Theories}
\author{A. Bonanno}
\address{
\baselineskip 14pt
Istituto di Astronomia, Universit\'a di Catania\\
Viale Andrea Doria 6, 95125 Catania, Italy\\
and\\
I.N.F.N. Sezione di Catania, Corso Italia 57, 95129 Catania}
\maketitle
\vspace{1 cm}
\baselineskip 14pt
\begin{abstract} 
\baselineskip 14pt       
A class of exact solutions of the field equations with higher 
derivative terms is presented when the matter field is
a pressureless null fluid plus a Maxwellian static  
electric component. It is found that the stable solutions are 
black holes in anti de Sitter 
background. The issue of the stability of the 
Cauchy horizon is discussed. 
\end{abstract}

\baselineskip 16pt
%
%
\section{Introduction}

Why should one wish to study the internal structure of a black hole?
Aside from sheer curiosity there is the intellectual challenge of
testing our present physical theories up to their extreme boundary of 
validity. 

In fact the spacetime outside a black hole is indeed a rather uninteresting
subject. Thanks to the no-hair theorem we know the 
late time structure of the spacetime after the star has radiated away 
all the asphericities \cite{abo-price}. Instead, the ultimate fate 
of the star that undergoes a gravitational collapse,
is still an open issue. 

The plausible, still unproven strong cosmic censorship conjecture states 
\cite{abo-penrose}
that the singularity at $r=0$ is spacelike. The spacetime
near the singularity can probably be described by the general mixmaster 
type solution  \cite{abo-BKL}. 

It is however puzzling that even an infinitesimally small amount of 
angular momentum generates the Kerr-Newman singularity 
which is instead timelike. 
In this case the situation is perhaps more dramatic due to the presence 
of a null hypersurface, the Cauchy horizon (CH), boundary of predictability
of the evolution of the fields. Like in the \RN solution, the 
existence of this  null hypersurface makes the existence of a timelike
singularity in a region beyond it physically irrelevant.

As it was pointed out by Penrose, \cite{abo-penrose} 
the Cauchy horizon is  also a surface of infinite blue-shift. A free-falling
observer crossing the Cauchy horizon will measure in-falling radiation to have
infinite energy density. When only ingoing radiation is present a weak
non-scalar singularity forms, which is classified as a whimper
singularity \cite{abo-ellis}. Whimper singularities are unstable to
perturbations that can transform them into  stronger, scalar,
singularities.

This scenario can be examined more closely by assuming spherical symmetry.
When an outflow crossing the Cauchy horizon is included in the analysis, 
the (Bondi) mass function is found to diverge exponentially\cite{abo-poi}
at late advanced times $v$
\be\label{abo-div}
m \sim e^{\kappa v}
\ee
where $\kappa$ is the surface gravity of the inner horizon located at 
$v=+\infty$. The only non-vanishing component of the Weyl tensor, 
the invariant $\Psi_2$, is proportional to the mass
function and a scalar curvature singularity forms 
along the Cauchy horizon \cite{abo-ori,abo-abo1}. 
This phenomenon, named ``mass inflation", has become a topic of 
increasing interest and lot of effort is devoted 
to study this subject\cite{abo-con}.

The \RN black hole shares the same causal
structure as the Kerr-Newman black hole, so it is not surprising 
that the basic  
picture derived in spherical symmetry should be qualitatively the same for 
the more general black hole. Some investigations suggest that the general 
scenario derived for spherical symmetry does not change dramatically 
in a non-spherical black hole \cite{abo-ori2,abo-crc,abo-abo2}.
In fact one key result of the mass inflation picture is that the
rate of the exponential divergence of the mass function is entirely 
determined by the surface gravity which is constant along the
generators of the Cauchy horizon. 

The relevant question is then to understand how much the classical 
description of the spacetime near the singularity changes when quantum 
effects are taken into account. In particular, would quantum 
effects enforce the strong censorship conjecture? 
A first analysis as been already performed in \cite{abo-wer2} where the 
semi classical 
approach has been used to show that the classical picture remains valid 
up to few Plank lengths from the CH. The mass function at that 
``time'' has already grown up to the mass of the observable Universe!

In fact, the classical description of the black hole interior 
is simplified by causality: behind
the event horizon the coordinate $r$ is timelike, so a descent into a black
hole is a progression in time. The evolution down to any particular radius is
only influenced by the initial data at larger radii. 
For this reason, a mean field, semi-classical 
description of quantum effects cannot drastically change the 
classical picture.  This scenario can instead change in two-dimensional 
models, as shown by \cite{abo-pom}.

However, there is no reason why one should assume that a perturbative
calculation can be applied when the curvature has reached planckian level. 
Only in a framework of a quantum gravity calculation one could hope 
to address this issue. 

In recent years it has become clear that a very  convenient way to
quantise gravity is to describe it as a quantum effective field
theory \cite{abo-adler,abo-don,abo-weinberg}.
As a result one is lead to consider a more general action that is 
a functional of any geometrical invariant $\Re$ which can be 
constructed from the principle of general covariance, $\Re=\{  
R, R_{\alpha\beta}R^{\alpha\beta}, C_{\alpha\beta\gamma\delta} 
C^{\alpha\beta\gamma\delta}, ...\}$, and of the matter field $\phi$ and 
its derivatives 
\be\label{abo-azi}
S[\phi,g]=\int d^4 x \sqrt{-g}{\cal L}(\Re,\;\phi,\;\nabla^2\Re, 
\nabla^2\phi, \;\nabla^{2j}\Re,\;\nabla^{2j}\phi,...,)
\ee
This new theory is analogous to the Euler-Heisenberg Lagrangian,
low-energy effective theory of the QED.
The important point is that both QED and the Euler-Heisenberg
theory represent {\it the same theory} at different scale
lengths. One would not include those new operators
in the QED Lagrangian since they are suppressed as
inverse power of the ultraviolet cutoff in the infrared region showing their
``irrelevant" character in this case. 
They can become instead significant
if we move the original cut-off of the theory in the infrared
region, where the theory must describe phenomena at much 
lower energy scales. 

This situation is similar in gravity. 
Due to the smallness of the ratio between the Newton constant and
the Fermi constant, modifications of Einstein's theory as in 
eq. (\ref{abo-azi}),
are experimentally  indistinguishable from
the standard theory  at ordinary energy scales \cite{abo-stelle}. 
However at higher energy scales it can alter the physics 
content of the theory in a significant way.
They might eventually destroy the unitarity of the S-matrix,
signalling that a new physics sets in at some energy scale.
The theory is perfectly well defined below that threshold,
and it makes precise predictions. 
This happens also in the so-called renormalizable
theories. The Standard Model, for instance, has to be
regarded as an effective field theory which breaks down at the TeV scale,
where the Higgs would enter in a strongly self-coupled phase. 

Higher-derivative  gravity theories arise also 
through the coupling of a quantised field with the classical 
background geometry \cite{abo-birrel} through the renormalization 
process. They are  
general functions of $\Re$ and are needed to cancel the  
ultraviolet divergences for any given order in perturbation theory. 

However, the standard perturbative
approach based on the scaling property of the Green's
functions under a rescaling of the metric 
\cite{abo-nelson} handles only a finite number of operators, those
which are important for the ultraviolet fixed point. 
In this way one cannot follow the evolution of  the irrelevant, 
i.e. non-renormalizable operators. 
A coupling is irrelevant, marginal
or relevant if it, respectively, 
it gets smaller, it does not run,  it grows 
when the cut-off is lowered from the ultraviolet 
towards the infrared \cite{abo-exp}. 
The irrelevant operators, even if they are not present in the bare
Lagrangian, mix their evolution with the renormalised trajectory 
of the relevant couplings. Although near the Gaussian ultraviolet fixed 
point their running is suppressed, they can behave in a quite 
different manner in other scaling region and they can eventually 
drive the system in a different continuum limit.

It is therefore important to trace the evolution of
all the coupling constants generated by the renormalization
procedure in order to study the phase structure of the theory. 
In this way one recovers the predictability power of the theory 
by retaining only few relevant operators at a given fixed point. 
If for example a new scale other than the cut-off 
is present in the problem, 
the scaling laws change near that scale and the
standard ultraviolet relevant interactions might not be sufficient
to describe the physics at a crossover between ultraviolet 
and infrared. 

The renormalization group approach used in Statistical Mechanics
is the best tool to study problems where many scales are
coupled together \cite{abo-wilson}. 
A powerful way of obtaining a differential form of the RG transformation
has been formulated by Wegner and Houghton \cite{abo-wegner}. Starting
from a bare action $S_k$ at the cut-off $k$ one first calculates 
$S_{k-\Delta k}$ in 
\be\label{abo-azi1}
e^{-S_{k-\Delta k}[\phi]}=\int {\cal D}[\psi]e^{-S_k[\phi+\psi]}
\ee
by using the loop expansion, where $\psi$ and $\phi$ 
respectively have non-zero Fourier 
components only in the momentum shells $k-\Delta k<p\leq k$ and 
$p\leq k-\Delta k$. The differential RG transformation 
is obtained by taking the limit of an infinitesimal shell
$\Delta k/ k \rightarrow \delta k / k$. 
The higher loop contributions in Eq. (\ref{abo-azi1}) are suppressed
as powers of $\Delta k/ k$ in the limit $\Delta k \rightarrow 0$ 
for finite $k$
and an exact, non-perturbative, one-loop RG equation is obtained   
\be\label{abo-wh}
k{d S_k[\phi]\over dk}=-{1\over 2}\langle \ln {\delta^2 S[\phi]
\over\delta\phi^2} \rangle +
{\langle {\delta S[\phi]\over \delta \phi} 
\Bigl ({\delta ^2 S[\phi]\over \delta \phi^2}\Bigr)^{-1}
{\delta S[\phi]\over \delta \phi}\rangle}
\ee
where the brackets indicates the sum over the Fourier components 
within the shell. This functional equation rules the evolution of all 
the interaction terms that are generated in the renormalization
procedure. It has been applied in \cite{abo-abo3} to discuss the 
evolution of higher-derivative (HD) operators generated by the inflation field. 
In particular it has been shown that even if they are not present 
at the cut-off - which has been chosen below the Planck scale! -
they show up above the mass threshold and influence the renormalised
flow. 
 
A consistent quantum gravity 
calculation in the framework of the effective theories,
must be performed by including the HD operators from the 
beginning \cite{abo-weinberg}. 
Within a perturbative (weak field) calculation one can use the 
Einstein-Hilbert truncation \cite{abo-reu}, 
but in a strongly non perturbative 
situation one should consider a more general Lagrangian  and, with the
help of the renormalization group equations, look for the
consistent field configurations that effectively 
dominate the path integral at that scale of energy.  

The first step towards this goal is to understand
the tree-level structure of the spacetime in the interior
of a black hole, when higher-derivative operators are 
included from the beginning. 

Those new operators can produce new istantons configuration 
in the path integral formulation of quantum gravity, with 
stronger weight than the solutions of the standard Einstein theory.
If one wants to perform a saddle point evaluation of the path integral 
the first step is to determine the stationary points of the 
(Euclidean) action. Expansion around those saddle points are performed
by writing
\be\label{abo-eee}
\gamma_{\mu\nu}=g_{\mu\nu}+h_{\mu\nu}
\ee
where, as usual, treating $g_{\mu\nu}$ as a background field and
$h_{\mu\nu}$ as a quantum field one generates the usual perturbation
expansion that can be expressed in terms of Feynman diagrams.
The strong assumption that is usually believed, is that vacuum state
for the renormalised system is still given by the $g_{\mu\nu}$
field configuration. Is this assumption still valid in the renormalised
system when new higher dimensional operators are generated by the
renormalization process?
If not, we have chosen a wrong, unstable, saddle point.
Strominger has shown \cite{abo-str} that for quadratic gravity, the Minkowsky 
solution is the only one having zero ADM energy. However that is
not necessarily true for a more general non-linear  gravity theory.
As it has recently been pointed out by \cite{abo-ovrut} where a
a completely new vacuum structure has appeared already for $R+R^3$ 
a Lagrangian. 

In this work we shall reduce the original HD theory to the standard
second order form by means of the method outlined in 
\cite{abo-ovrut,abo-guido}. We shall obtain an equivalent (locally
isomorphic) second order theory with an additional, non-geometric
degree of freedom represented by a non-ghostlike scalar field. 
We call ``vacuum" a solution which is stable with 
respect to the elementary  
excitation of this new field. 
We show that for a general non-linear Lagrangian  
the theory has non-trivial vacuum solutions 
representing BH embedded in Anti de Sitter spacetime.

We shall also comment on the 
stability of the Cauchy horizon in those solutions.

%
%

\section{basic equations}
Let us consider the following action
\be\label{abo-11}
S = \int d^4x \sqrt{-g} (l_g (R) + l_m({\cal A},g))
\ee
where $l_g(R)$ is a non-linear function of the scalar curvature $R$ which we
shall suppose to be an analytic function of  $R$\
\be\label{abo-12}
l_g(R) = \sum_n^{\infty} {\epsilon_n \over n!}R^n
\ee
$\epsilon_n$ are the coupling constants of the generic 
power of $R$ and $l_m({\cal A})$ is the Lagrangian density of 
the matter field ${\cal A}$. 
We use the signature (-+++) and we set $c=\hbar=8\pi G$.
With these units 
$\epsilon_0=-\Lambda$ 
corresponds to the vacuum energy, being $\Lambda$ the cosmological constant
and $\epsilon_1 = 1/2$. 
The gravitational field equations for such a system are 
the following fourth order differential equations 
\be\label{abo-13}
{\delta \over \delta g_{\mu\nu}}[\sqrt{-g}l_g(R)] = l'_g(R)R_{\mu\nu}
-{1\over 2}l_g(R)g_{\mu\nu}-\nabla_\mu\nabla_\nu l'_g(R)+
g_{\mu\nu}\Box l'_g(R) = T_{\mu\nu}({\cal A},g) 
\ee
where $T_{\mu\nu}$  
is the stress-energy tensor of the matter field which 
obeys $\nabla_{\mu}T^{\mu\nu}=0$.
 
In order to analyse the above theory it is convenient \cite{abo-ovrut} to introduce 
an auxiliary field $\psi$ so that the action now reads 
\be\label{abo-14}
S=\int d^4 x \sqrt{-g}\Big ( l'_g(R)(R-\psi)-l_g(\psi)+l_m(A,g)\Big )
\ee
with the equation of motion 
\be\label{abo-15}
l_g''(\psi)(R-\psi) = 0. 
\ee
The new action, written with the help of the auxiliary field, is
(at the classical level) equivalent to the original theory provided that 
we are not at the critical points $l_g''(R) = 0$, but it is not of the
second order form.  
In order to reduce the action (\ref{abo-15})
in a canonical second order form, one introduces a pair of new variables
$(\bar{g}_{\mu\nu}, \omega)$ related to $g_{\mu\nu}$ by
\be\label{abo-16}
e^{\omega} = l_g'(R);\hspace{2 cm} 
\bar{g}_{\mu\nu}=e^{\omega}g_{\mu\nu}
\ee
where $l_g'>0$ is assumed   
and $\Psi(\omega)$ becomes a solution (not necessarily unique), 
of 
\be\label{abo-edd}
l_g'(\Psi(\omega))-e^\omega = 0.
\ee
This transformation has already been used in the 
literature, and it has been generalised to the case of
Lagrangians depending  on $\nabla^{2k}R$ (see 
\cite{abo-guido} for a general review on the subject).
Under the above conformal transformation 
the Ricci scalar density transforms like 
\be\label{abo-18}
\sqrt{-g}R=\sqrt{-\bar{g}}e^{-\omega}
\Big (\bar{R}-{3\over 2}
(\bar{\nabla}\omega)^2-{3}\bar{\nabla}^2\omega \Big ).
\ee
By dropping a total derivative term, the action has been reduced
to the canonical form   
\be\label{abo-19}
S=\int d^4x\sqrt{-\bar{g}}\Big (\bar{R}-{3\over 2}\bar{g}^{\mu\nu}\partial_\mu\omega
\partial_\nu\omega -2 V(\omega) +e^{-2\omega}l_m({\cal A},
{e^{-\omega}\bar{g}})\Big )
\ee
where the potential term 
\be\label{abo-110}
V(\omega)={1\over 2}{\rm e}^{-2\omega}\Big (
{\rm e}^{\omega} \psi(\omega)-l_g(\psi(\omega))\Big ).
\ee
depends on the specific original higher-derivative theory. The field equations 
thus read    
\be\label{abo-111}
\bar{G}_{\mu\nu} = t_{\mu\nu}+\bar{T}_{\mu\nu}
\ee
where $\bar{G}_{\mu\nu}$ is the Einstein tensor of the reduced theory
\be
\bar{G}_{\mu\nu} = \bar{R}_{\mu\nu}-{1\over 2}\bar{g}_{\mu\nu}\bar{R} 
\ee
$t_{\mu\nu}$ is the stress energy-tensor of the scalar field
\be\label{abo-scat}
t_{\mu\nu}={3\over 2}\Big [\partial_\mu\omega\partial_\nu\omega-{1\over 2}
\bar{g}_{\mu\nu}\bar{g}^{\alpha\beta}\partial_\alpha\omega\partial_\beta\omega\Big ]
-\bar{g}_{\mu\nu} V(\omega)
\ee
and $\bar{T}_{\mu\nu}$ an effective stress-energy tensor generated by
the interaction between the original 
matter field and the non-geometric gravitational new degrees of freedom
represented by the scalar $\omega$,
\be\label{abo-ints}
\bar{T}_{\mu\nu} = e^{-\omega}T_{\mu\nu}
({\cal A},e^{-\omega}\bar{g}).
\ee
One should also note that in general the two terms on the left
hand side of the equation (\ref{abo-111}) are not separately conserved,
but it holds instead the conservation law $\nabla_\mu\bar{G}^{\mu\nu}=0$.
We stress that the spin-0 field introduced with the help of
eq.(\ref{abo-16}) has physical meaning since it corresponds
to an additional degree of freedom already present in the starting
non-linear Lagrangian. Therefore the advantage of having 
used this approach is that this new degree of freedom has been
now made explicit in the reduced Lagrangian. 
 
Since we are interested in studying the structure of the CH in the presence of HD 
terms, we consider a system where the gravitational field is coupled with 
a pressureless null fluid plus an electromagnetic contribution coming from
a static electric field generated by a charge of 
strength $e$. The stress-tensor for the matter field
is therefore given by
\be\label{abo-ene}
T_{\mu\nu} = \rho l_\mu l_\nu + E_{\mu\nu}
\ee
where $E_{\mu}^{\nu}= e^2/8\pi r^4{\rm diag}(1,1,-1,-1)$ is the Maxwellian component of the
electric field,  $\rho$ is the energy density of the radiation and 
$l_\mu$ is a null vector tangent 
to the radial null geodesic. In the following we shall 
specify our calculations in the case of spherical 
symmetry. One can introduce a null tetrad
$\{ l_{\mu},n_{\mu},m_{\mu},\bar{m}_\mu \}$  
with $l_\mu n^\mu = -1 = -m_\mu \bar{m}^\mu$. 
The metric tensor then reads 
\be\label{abo-ene2}
g_{\mu\nu}=2
\Big [\bar{m}_{(\mu}m_{\nu)}-l_{(\mu}n_{\nu)}\Big ].
\ee
and 
\be\label{abo-ene1}
E_{\mu\nu} = {e^2\over 4\pi r^4}[l_{(\mu}n_{\nu)}+\bar{m}_{(\mu}m_{\nu)}]
\ee
In our case the matter Lagrangian is conformally invariant,
and the equation of motion for the matter field completely decouple from
those for the scalar field.
We analyse the case of $\omega ={\rm const.}$ and the solutions of the
equation of motion correspond to constant field configuration 
that are extrema for the potential
\be\label{abo-113}
{d V\over d\omega} = 0
\ee
The vacuum case -no matter field- has been analysed in \cite{abo-ovrut}. 
In our case we see that      
the conformal transformation in (\ref{abo-16}) generates a conformal
rescaling of the null tetrad in  (\ref{abo-ene2}) with $l_\mu \rightarrow 
\tilde{l}_\mu=\exp(\omega/2)l_\mu$. It is then easily seen that 
the stress-energy tensor in the reduced theory in eq.(\ref{abo-ints})
has the same functional form of that in the original theory, 
with the ``renormalised'' energy density and charge
\be\label{abo-114}
\rho\rightarrow \bar{\rho}= \exp(-2\omega)\rho \hspace{1 cm}
e^2\rightarrow \bar{e}^2=\exp(-2\omega) e^2.
\ee
The solutions of the new field equations (\ref{abo-111}) can therefore be classified
by looking at the minima of the potential term (\ref{abo-110}).  
Even if in the original theory a cosmological term is not present,
the solutions  of the reduced theory are given by 
a Vaydia like spacetime in a de Sitter (dS) or Anti-de Sitter
(AdS) background, depending on the value of the potential at minimum. 
In particular, one can use the Eddington-Finkelstein coordinates so that an 
explicit form of the metric
element in the reduced theory reads
\be\label{abo-36}
d\bar{s}^2=dv(2dr-fdv)+r^2 d\Omega^2.   
\ee
where 
\be\label{abo-37}
f=1-{2m(v)\over  r}+{\bar{e}^2\over r^2}-
{\Lambda \over 3}r^2
\ee
is the so called mass function, 
$\Lambda=V(\omega_c)$  and $\omega_c$ is the location of any 
extrema of the potential. In a local chart adapted to the inner horizon
the Cauchy horizon is located at $v=+\infty$, and $f=0$.
If we compare this metric with the solution we would
have gotten in the Einstein gravity, we note that,
according to eq.(\ref{abo-114}) the charge and the energy densities 
are screened or anti-screened depending on the sign of $\omega_c$. 
It should be also noted that, while in standard gravity one
has always AdS solution of the kind (\ref{abo-36}) in our case 
the above metric is a solution of the field equations only for
special values of the cosmological constant.

The global spacetime structure of those solution is well 
known. An updated review 
can be found in the contribution by Chris Chambers in these proceedings, 
while recent results on topological (AdS) 
black holes can be found in the lecture
of R. Mann \cite{abo-cama}. See also there for the figures
displaying the conformal structure of the two spacetimes.

Here we shall simply stress that the black hole 
spacetime in both dS and AdS solutions may have
Cauchy horizons depending on the roots of the equation
\be\label{abo-rot}
f(r)=0 
\ee 
If the dS case is considered eq.(\ref{abo-rot})
may have three positive roots and the spacetime has therefore
a cosmological horizon besides an inner and an event horizon. 
Thanks to the work of Brady and Poisson 
\cite{abo-brapo} we know 
the Cauchy horizon is stable if the surface gravity of the cosmological
horizon is greater than the surface gravity of the inner horizon
\be\label{abo-sta}
\kappa_{CS}>\kappa_{CH}
\ee
The surface gravity of the horizon in the 
original theory is obtained by a simple scale transformation
(from the definition of surface gravity), and one can conclude 
that the stability criteria 
(\ref{abo-sta}) holds in the original theory as well.

However, a black hole in dS background does not represent   
a vacuum solution
since the field sets on a maximum of the potential. 
The stable vacuum solutions
for the excitations of the $\omega$ field 
are instead given by black holes in AdS background and for those solutions the 
Cauchy horizon is a subtle question. The spatial
infinity is now timelike, the radiative falloff at late times
is not well understood \cite{abo-29}. For a power law 
decay of the kind 
\be\label{abo-deca1}
\delta(v)=1/(\alpha v)^p
\ee 
with $p\ge 3$ and $\alpha$ has to be introduced 
on dimensional grounds, the Cauchy 
horizon is unstable. In fact in general the energy 
density $\rho_{obs}$ measured by an observer that crosses the Cauchy horizon
would diverge as
\be\label{abo-dic}
\rho_{obs}\sim {\dot\delta} \exp({2\kappa_{CH} v}) 
\ee
thus signalling the instability of the Cauchy horizon.
The CH in the original theory would also be
unstable since the conformal transformation in 
eq. (\ref{abo-16}) is regular.
However for an exponential decay rate of the kind 
\be
\delta \sim \exp(-2\alpha v)
\ee
one would not see any divergence for 
\be
\alpha > \kappa_{CH}
\ee
In a more realistic situation, one should use the eq. 
(\ref{abo-deca1}) to mimic the radiation falling into 
the event horizon. Outside the event horizon
there is no reason to suppose that the background being
AdS. Our effective ``cosmological" constant is therefore
dynamically generated during the gravitational collapse for the 
presence of the quantum fluctuations when the curvature
has reached planckian level. The initial data
coming from larger radii is still given by the 
classical evolution and therefore the Cauchy horizon
would still be unstable even in the presence of a
negative (or positive) $\Lambda$ term in the metric. 


\section{$R^2$, $R^3+R^2$ gravity}
Let us analyse some cases in  detail.
If we consider $R^2$ gravity, the density Lagrangian of the gravitational 
field reads
\be\label{abo-20}
l_g(R) = \epsilon_0+\epsilon_1 R+{1\over 2} \epsilon_2 R^2.
\ee
then, eq. (\ref{abo-edd}) can be inverted to yield $\psi=\ln(1/2+
\epsilon_2\psi)$ provided $\psi >-1/2\epsilon_2$. In particular
the potential term becomes
\be\label{abo-21}
V(\omega) = {1\over 2\epsilon_2}(1-{\epsilon_1} e^{-\omega})^2-
e^{-2\omega}\epsilon_0
\ee
We see that for a positive 
cosmological constant in the original theory the potential term is not bounded 
from below and there are no stable solutions of the quadratic
gravity theory. 
In the case of a negative cosmological constant, the potential term has only one minimum  for
$\omega_c=-\ln 2$ and we find the BH solutions given by eq.(\ref{abo-37}).
We note that the "renormalised" charge and energy density  
have now an increased strength of a factor $4$ according to (\ref{abo-114}). 

Another interesting case is that of $R^3$ gravity where we suppose that 
the original Lagrangian has the following functional form 
\be\label{abo-22}
l_g(R) = \epsilon_1 R + {1\over 2}\epsilon_2 R^2 +{1\over 3!}\epsilon_3 R^3
\ee
in this case we can invert the relation (\ref{abo-edd}) in order to obtain the
two solutions 
\be\label{abo-23}
\psi{\pm}(\omega)={1\over \epsilon_3}\Big (-\epsilon_2\pm\sqrt{\epsilon_2^2
+2(e^{\omega}-\epsilon_1)\epsilon_3)}\Big )
\ee
in the two branches 
$\psi>-\epsilon_2/\epsilon_3$ and $\psi<-\epsilon_2/\epsilon_3$ and for the reality
condition we also consider $\omega>\omega_{*}=\ln(1/2-\epsilon_2^2/\epsilon_3)$. 
The potential term is therefore given by   
\be\label{24}
V(\psi_{\pm}(\omega))={\epsilon_2\over 2}\psi_{\pm}^2(\omega) e^{-2\omega}
\Big (1+{2\epsilon_3\over 3\epsilon_2}\psi(\omega)\Big )
\ee

In the first case the potential $V(\psi_{+}(\omega)$ has one minimum
for $\omega_c=-\ln 2$ with $V(\omega_c)=0$ and a local maximum at another 
$\omega_c<0$ with $V(\omega_c)>0$. 
Therefore in the $\psi_{+}$ branch we can conclude that the solution of the 
field
equations is given by a Vaidya like metric in a dS background.
In the $\psi_{-}$ branch we have only one local minimum that 
corresponds to a negative value of the potential, and
therefore the metric is still a Vaidya like, but in AdS background.

Other interesting cases are given by  Lagrangian like
$l_g(R)=R+bR\ln(a+R)$, with $a$ and $b$ constants,
which would arise from a one-loop
contribution in an interacting stress-energy tensor. In any 
case one can use the above outlined procedure, and derive
the explicit expression of the dS or AdS black hole solution. 

The use of the conformal transformation is not the only method to obtain
the solution of the original theory. 
There is also a direct procedure which is of course equivalent.
Let us consider the field equations given in (\ref{abo-13}) and set 
\be\label{abo-31}
R_{\mu\nu}=C(R)\Big(\rho l_{\mu}l_\nu+E_{\mu\nu} \Big)
+{1\over 4}R g_{\mu\nu} 
\ee
with constant Ricci scalar $R$. The following statement is true: 

{\it 
The solutions of the field equations in (\ref{abo-13})
with the stress energy-tensor in (\ref{abo-ene}) and the 
Ricci tensor of the form (\ref{abo-31}) have}
\bea\label{abo-32}
&&R l_g'(R)-2 l_g(R) = 0\\[2mm]
&&l_g'(R)C(R)=1 .
\eea
The above conditions have to be considered as implicit 
relations that constrain the possible values of the constant $R$.
They can be deduced by direct substitution of (\ref{abo-31})
in (\ref{abo-13}) and by projection along $l^\mu n^\nu$ and $n^\mu n^\nu$ 
respectively.  One also notes that the first of the above relations
in (\ref{abo-31}) is equivalent, aside from a constant factor, to the extrema 
condition for the potential, and the second relation can be read off as
$C(R)=e^{-\omega}$ which is the first of the relations in
(\ref{abo-16}). This shows  the 
equivalence of the 
two approaches in obtaining solutions of the HD theory.
However the conformal transformation method is simpler and
elegant, while the direct method has the advantage that can also be
used when $l''_g(R)=0$.
The set of consistency equations in (\ref{abo-31})
can also be deduced by considering the following stress energy tensor 
\be\label{abo-34}
T_{\mu\nu}=C(R)^{-1}\rho l_{\mu}l_{\nu}+E_{\mu\nu}
\ee
and with the Ricci tensor given by 
\be\label{abo-35} 
R_{\mu\nu}=\rho l_{\mu}l_\nu+C(R)E_{\mu\nu} 
+{1\over 4}R g_{\mu\nu} .
\ee
The above relations can be used to investigate the full global 
content of the theory. Once eq.(\ref{abo-31}) has been solved,
one introduces locally the conformal frame where the new degrees
of freedom is made explicit. Then one can address the question
of local stability and mass of the field. 

\section{conclusions}
As we have stressed in the introduction, the non-linear 
modifications of the standard Einstein theory are needed for taking 
into account of the quantum effects when the Weyl curvature 
has reached Planckian levels.
We have seen that stable black hole solutions appears in these cases.
What is the possible role of these new solutions in the standard
mass inflation scenario? 
It is interesting to understand the structure of the
CH singularity in a more consistent calculation
where these new terms are dynamically generated during the
collapse. We have argued that the role of those new 
terms would not change the (unstable) character of the Cauchy 
horizon.
We therefore think that the mass inflation phenomenon 
also takes place in such a background. However since the
appearance of an effective $\Lambda$ term changes both
the location of the horizon and the surface gravity of the 
inner horizon, a detailed calculation must be performed in order to
address this issue.

\section{acknowledgements}
The author thanks the organiser of  "The Internal Structure of Black 
Holes and Spacetime Singularity" workshop, at Haifa for the very pleasant 
meeting and for the friendly hospitality at Technion. 
My warmest thanks go to Amos Ori, Lior Burko and Liz Youdim for having
organised this interesting meeting. I have personally benefited 
from discussions with Lior Burko, Chris Chambers, 
Valery Frolov,  
Robert Mann and Amos Ori whom I gratefully acknowledge.
I would also thank Daniela Recupero for  
careful reading the manuscript.


\begin{thebibliography}{99}
\bibitem{abo-exp} The correct definition of irrelevant and marginal 
operators should be given after linerization of the renormalization group 
equation around a fixed point. We are not interested in these technical 
issues.
\end{thebibliography}
\end{document}